\documentstyle[prl,aps]{revtex}
\input epsf
\input{psfig.tex}
\begin{document}

\title{Quintessence and variation of the fine structure constant in the CMBR}

\author{$^1$Greg Huey, $^2$Stephon Alexander, and $^2$Levon Pogosian}

\address{
$^1$Astronomy Unit, School of Mathematical Sciences, Queen Mary College, London E1 4NS,
United Kingdom.
\\$^2$Theoretical Physics, The Blackett Laboratory, Imperial
College, London SW7 2BZ, United Kingdom}

\wideabs{
%\twocolumn[
\maketitle

\begin{abstract}
\widetext We study dependence of the CMB temperature anisotropy spectrum
on the value of the fine structure constant $\alpha$ and the equation of state
of the dark energy component of the total density of the universe. We find that
bounds imposed on the variation of $\alpha$ from the analysis of currently
available CMB data sets can be significantly relaxed if one also allows
for a change in the equation of state.
\end{abstract}
\pacs{} }
%]

\narrowtext

\section{Introduction}
\label{introduction}

Current observations of type Ia supernovae \cite{SNIa} suggest that our Universe
is accelerating. This has lead many theorists to allow for the
existence of a mysterious dark energy
that permeates the universe and has negative pressure.
One example is a cosmological constant, $\Lambda$, with the equation of state
$w_{\Lambda} \equiv p_{\Lambda}/\rho_{\Lambda} = -1$.
More recently, it was suggested that the dark energy would not necessarily
have to be of constant density at all times.
The idea is to introduce a dynamical light scalar field $Q$, called Quintessence, with a
tracking potential $V(Q)$ chosen in such a way that $Q$ comes to dominate the expansion
of the Universe only recently. The equation of state
$w_{Q} \equiv p_{Q}/\rho_{Q}$ will now depend on the choice of $V_Q$ and will generally
be time-dependent. The current value of $w$ of dark energy is only loosely constrained:
$w \lesssim -0.6$ \cite{pel99}, however there
is hope that future experiments will improve the bounds \cite{weller01}.

Dirac was among first to suggest that fundamental constants, such as
the fine structure constant $\alpha\equiv e^2/\hbar c$, could vary with time \cite{dirac}.
The interest in varying constant theories has recently risen with the
increased popularity of models with both large and small extra dimensions
in which four-dimensional constants are no longer fundamental.
Additional motivation is provided by the fact that some of the
puzzles of Cosmology, such as the horizon, flatness and, arguably, other problems
as well, could be resolved if the speed of light was larger in the past
\cite{Moffat93,albrecht99}.

Experimental constraints on variation in $\alpha$ come from
atomic clock tests \cite{Demidov92,Prestage95}, measurements of isotope ratios
\cite{Sh76,oklo} and absorption spectra in distant quasars
\cite{savedoff56,webb}. While all laboratory and geophysical tests have so far failed
to see any indication of $\alpha$ varying at present epoch \cite{Varsh00},
the quasar data \cite{webb} has produced some evidence that the
fine structure constant could have been smaller in the past.

The imprint of a varying fine structure constant on the
cosmic microwave background radiation (CMBR) has been studied before \cite{Han99,Kap99}.
It is usually assumed that the value of $\alpha$ at the time of last-scattering
was different from its present value but that it did not change considerably throughout
the recombination epoch. It is also assumed that at any given time $\alpha$ was
the same everywhere in space\footnote{While we employ the same assumptions
in this work, we would like to stress that in many varying constant theories
the change in $\alpha$ comes from the dynamics of a time- and space-dependent
scalar field \cite{Haavard,Moffat,Dvali01}. Fluctuations of this scalar field could
potentially have a non-trivial effect on CMBR.}. A change in $\alpha$ at the
time of recombination would change the cross-section of Thomson scattering of
CMB photons and also would alter the energy levels of atoms. Thus,
the main effect of varying $\alpha$ comes from the change in the redshift of
the last scattering surface.

It was argued in \cite{Han99} and \cite{Kap99} that the next generation
of CMB experiments should be able to constrain the variation of $\alpha$ at redshifts
$z \sim 1000$ with an accuracy
$\Delta_{\alpha} \equiv (\alpha - \alpha_0)/\alpha_0 \sim 10^{-2}-10^{-3}$,
where $\alpha_0$ is the current value.
The likelihood of a varying $\alpha$ based on the recent CMB data \cite{boom,maxima,dasi}
was analysed in \cite{Ave00,Bat01,Ave01}. While in \cite{Ave00} and \cite{Bat01}
it was found that the data prefers a smaller value of $\alpha$ in the past,
the combined analysis of the most recent CMB data and the big bang nucleosynthesis
(BBN) constraints in \cite{Ave01} did not produce any
evidence for a varying $\alpha$ at more than 1$\sigma$ level.

In all previous studies, when looking at the effect of varying $\alpha$ on
CMB spectra, it was assumed that the vacuum energy of the universe is due to a
cosmological constant. Alternatively, the vacuum energy could be due to a quintessence
field. One might question if the constraints on the change in $\alpha$ would be different
if the variation in $\alpha$ was considered at same time with the variation in
the equation of state of the dark energy component.
While the change in the fine structure constant effectively changes the redshift
of the CMB last-scattering
surface, a change in the equation of state of quintessence changes
the conformal distance to a fixed redshift.
Thus, to some extent, changes in the CMBR anisotropy spectrum caused by
$\Delta _{\alpha }\neq 0$ can be compensated for by a change in
$w_{Q}\equiv p_{Q}/\rho _{Q}$.
We shall see that this indeed is the case.

We would like to emphasize that this work is an exposition of a degeneracy
- not an evaluation of experimental constraints.

This paper is organized as follows. In Section \ref{implement} we describe our
implementation of quintessence and varying $\alpha$. In Section \ref{degeneracy}
we discuss how we search for degeneracies in CMB spectra. The
results are presented in Section \ref{results} and we finish with a discussion in
Section \ref{outlook} of possible theoretical frameworks in which the quintessence
field and the variation in $\alpha$ may be inter-related.

\section{Quintessence, varying $\alpha$ and CMBR}
\label{implement}

There is an enormous variety of quintessence models, {\it i.e.} models containing a
dynamical scalar field which could drive the current accelerated expansion of the universe.
While any particular choice of a model, or even a class of models, still
remains a matter of personal taste, there is a relatively limited set
of properties relevant to the CMBR.
The effect of the $Q$-component on the CMBR spectra
is primarily due to the change in the conformal distance to the last scattering surface.
Somewhat less prominent is the role of the perturbations in the $Q$-component.
It was shown in \cite{Caldwell98} that the main effect of including the perturbations
is on very large scales due to the integrated Sachs-Wolfe effect (ISW).
For completeness, in this work we do take into account the fluctuations in the pressure and energy density
of the quintessence.

We have assumed that the quintessence field (or the $Q$-component)
couples to other particle species only gravitationally.
The evolution of the energy density
and the pressure of the $Q$-component as well as their perturbations is
completely specified by the equation of state (EOS), $w(t)= p_Q/\rho_Q$,
which is generally time-dependent. Equivalently, one could start with the potential
$V_Q$ of the $Q$-field and deduce the EOS from it. However, many different potentials
can lead to the same EOS.

We will limit ourselves to models in which the EOS
of the $Q$-component remains effectively constant between the
time of recombination and today. The reason for this restriction is simply
the fact that none of the specific quintessence models appears to be more attractive
than others. We therefore take the simplest case. The effects of
several time-dependent EOS were examined in \cite{Caldwell98}. Predictions of
some specific models were also studied in \cite{Huey99,Bean01}.

The effect of varying fine structure constant $\alpha$ on the CMBR comes from the changes
in the differential optical depth $\dot{\tau}$ of photons during the
time of recombination. $\dot{\tau}$ can be written as
\begin{equation}
\dot{\tau}=x_e n c \sigma_T \, ,
\end{equation}
where $x_e$ is the ionization fraction, $n$ is the electron number density and
$\sigma_T$ is the Thomson scattering cross-section.
The dependence of $\sigma_T$ on $\alpha$ is well known:
\begin{equation}
\sigma_T = {8 \pi \alpha^2 \hbar^2 \over 3 m_e^2 c^2} \,
\end{equation}
where $m_e$ is the electron mass. The ionization fraction $x_e$
depends on $\alpha$ through the binding energy of hydrogen as well
as through the change in the recombination rates.
The correct procedure for accounting for these two effects is described in
\cite{Han99,Kap99}. We have closely followed the discussion in \cite{Han99,Kap99}
when incorporating the effects of varying $\alpha$ into CMBFAST \cite{cmbfast}.

We will be calculating the angular power spectrum $C_l$ of the
CMB temperature anisotropy defined as following:
\begin{equation}
C_l = {1\over 2l+1}\sum_{m=-l}^{l} \langle a^*_{lm} a_{lm}\rangle \, ,
\end{equation}
where
\begin{equation}
a_{lm} = \int d\hat{n} \,
Y^*_{lm}(\hat{n}) \left({T(\hat{n})-\bar{T} \over \bar{T}}\right) \, ,
\end{equation}
where $T(\hat{n})$ is the CMBR temperature in a certain direction on the sky and
$\bar{T}$ is the average temperature.

\section{Searching for degeneracies}
\label{degeneracy}

The CMB anisotropy spectrum is computed by a version of CMBFAST \cite{cmbfast},
modified for simultaneous quintessence (including perturbations in
quintessence) and variable fine structure constant. We have only considered
flat models ($\Omega_{total}=1$) with adiabatic initial conditions.
The parameter space consists of
($\Omega_m, w_Q, \Delta_{\alpha}, h, \Omega_B h^2, n_s, N $),
where ($\Omega_m$ is the ratio of the cold dark
matter energy density to the critical density, $w_Q$ is the
quintessence equation of state,
 $\Delta_{\alpha}\equiv(\alpha-\alpha_0)/(\alpha_0)$ is the
fractional change
in the fine structure constant, $h$ is the Hubble constant in
units of 100 km s${}^{-1}$Mpc${}^{-1}$, $\Omega_B h^2$ is the baryon density,
$n_s$ is the scalar spectral index and $N$ is the overall
normalization of the spectrum. The restriction to flat geometry
implies that $\Omega_Q = 1- \Omega_m $.
Each point in this parameter space has a CMB
anisotropy spectrum associated with it.
Two points in parameter space are considered degenerate if
their associated CMB spectra are indistinguishable.
The degeneracy of the parameter space is surveyed by
picking a point in that space to be the fiducial model,
and then comparing its CMB spectrum with that of
another point.
To illustrate the degeneracy in the $ (w_Q, \Delta_{\alpha}) $
plane, these parameters were gridded. A fiducial model was picked
and its spectrum was compared to the least distinguishable spectrum
of each grid point. The parameters $\Omega_m$
and $\Omega_Q$ were held fixed, while $h, \Omega_B h^2, n_s, N $
were allowed to vary to find the model least distinguishable
from the fiducial model. It should
be emphasized that the parameters of the fiducial model are chosen to
suit illustrative purposes, and are not always related
to experimental observations. Our results are an exposition of a degeneracy
in parameter space - not an evaluation of experimental constraints.

The presence of degeneracy sensitively
depends on the criteria one uses to determine distinguishability
of CMB spectra. A real CMB anisotropy experiment would be limited
in the following ways: a finite beam width would imply a minimum
scale resolution, which we approximate as a simple truncation of
the spectrum above a specific $ \ell _{max} $. Normally the
results of the experiment would be analyzed as independent 'bins' -
effectively a collective range of $ C_{\ell }$'s. We consider
a bin size of 1 multipole in our runs, as this illustrates the degeneracy
present in an optimistic future experiment. Finally, any real experiment will
have some level of error above cosmic variance - due to incomplete
sky coverage, instrumentation noise,
non-CMB sources in the sky, etc. Again intending to demonstrate the
optimistic limit, we take the error at each $ C_{\ell }$ to be
 $5\% $ plus cosmic variance.
To address the issue of different levels of degeneracy, some runs
are done with CMB spectra being computed and compared out to a maximum
multipole of $ \ell _{max}=900 $ - this captures the large-angle
plateau and acoustic peak structure, but not the damping scale. Alternatively,
some runs are done with $ \ell _{max}=1500 $ which additionally
captures the damping scale.

Note that only the scalar portion
of the spectrum is considered here. The addition of a tensor component
would not diminish the degeneracy - instead, as we discuss below,
it might qualitatively increase it.

The chi-square from the comparison
of the CMB spectrum of each point on the grid with the fiducial model
is used to determine how distinguishable the points are. In the plots,
solid curves mark the contours of $68.3\%$, $95.5\%$ and $99.7\%$
likelihood distinguishability. That is, points on the outer contour
produce CMB spectra such that one can say with $ 99.7\% $ confidence
that these spectra are not produced by the fiducial model.

\section{Results}
\label{results}

We have found that effects of changing the fine structure constant and
varying the equation of state of quintessence are to a large extent degenerate.
This degeneracy arises because it is possible to compensate for the
change in the redshift of last-scattering ($ \Delta _{\alpha } $) by a change
in the conformal distance to a given redshift ($ w_{Q} $) - the
quantity that must remain fixed is the angle on our sky subtended
by the sound horizon at last-scattering (ie: the angular scale of
the first Doppler peak). In addition to changing the
redshift of the last-scattering, a change in $ \Delta _{\alpha } $
also changes the thickness of the last-scattering surface.
Anisotropies on scales shorter than this thickness destructively
interfere when projected onto the sky and, as a result, the CMB power spectrum
is suppressed below a certain scale.
Also, at very small scales perturbations in the primordial plasma are washed out
due to the imperfect coupling of baryons and photons - the Silk damping, which
further suppresses the CMBR anisotropy spectrum.

\begin{figure}
\psfig{file=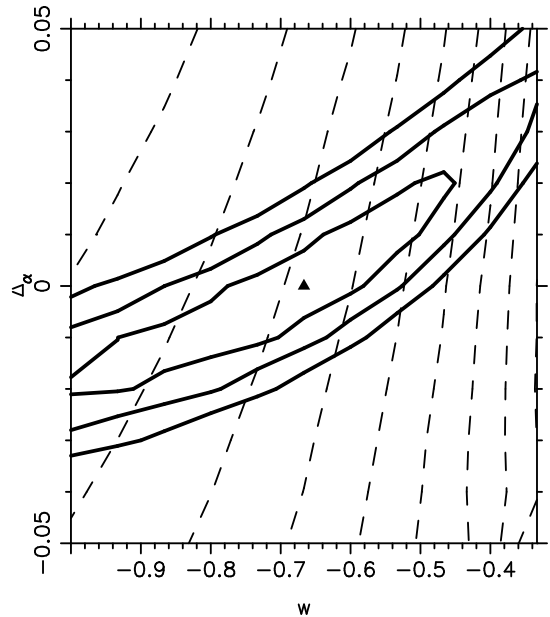,height=8.5cm}
\caption{Degeneracy around a quintessence fiducial
model (marked with a solid triangle) in the
$\left( w_{Q},\Delta _{\alpha }\right)$
plane with $w_{Q}=-2/3$, $\Delta _{\alpha }=0$.
The models are compared to the fiducial model at each $ C_{\ell }$
out to $\ell _{max}=900$. The solid curves are
$68.3\%$, $95.5\%$ and $99.7\%$ likelihood bounds,
and the dashed lines are contours of constant baryon density.
The fiducial model has $\Omega_B h^2 = 0.0200$
and $\Omega_B h^2$ changes by 0.0007
between the dashed contours.}
\label{fig_degen_qFM_900}
\end{figure}

The extent of the degeneracy depends on what
criteria one uses to determine which CMB spectra are in principle
distinguishable and which are not.
If one measures the anisotropy only
on scales larger than the damping scale ($ \ell \sim 900 $),
compares each $ C_{\ell } $ individually (rather than binning them),
and considers a modest ($ 5\% $) error in addition to cosmic variance,
then one does find the degeneracy, as shown in Fig.~\ref{fig_degen_qFM_900}.
For the fiducial model (marked with a triangle on the plot) we have chosen $w_Q=-2/3$
and $\Delta_{\alpha}=0$. Similarly, we have tried using $ w_Q=-1$ for the fiducial
model and found that the degree of the degeneracy remains unchanged.
The rest of parameters in the fiducial model were chosen to be $\Omega _{total}=1$,
$\Omega _{m}=0.3$, $h=0.65$, $\Omega _{B}h^{2}=0.02$ and $n_{s}=1$.

\begin{figure}
\psfig{file=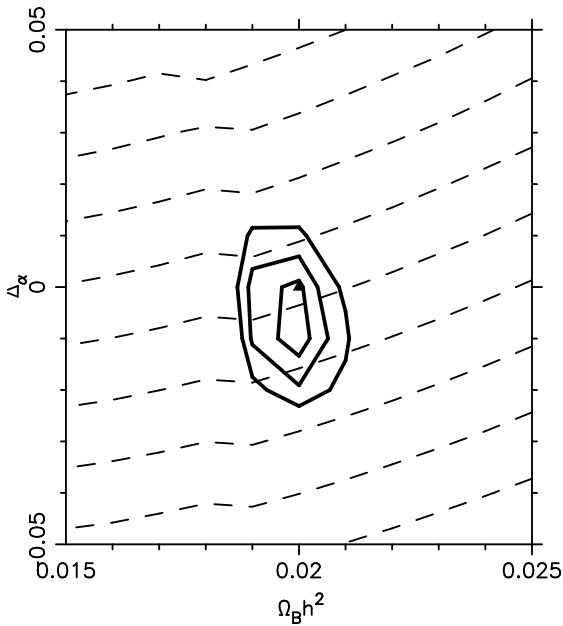,height=8.5cm}
\caption{Likelihood contours in the
$\left( \Omega _{B}h^{2},\Delta _{\alpha }\right)$
plane around a quintessence fiducial model (marked by a star)
with $w_{Q}=-2/3$,
$\Omega _{B}h^{2}=0.02$, $\Delta _{\alpha }=0$.
The models are compared to the fiducial model (marked with a solid triangle) at each
$C_{\ell }$
out to $ \ell _{max}=900$. The solid curves are
 $68.3\%$, $95.5\%$ and $99.7\%$ likelihood bounds, and the dashed lines are contours
of constant scalar spectral index. The value of the spectral index changes by
$0.031$ between each dashed line. Note the absence of the degenerate
direction.}
\label{fig_baryoncontrol_qFM_900}
\end{figure}

\begin{figure}
\psfig{file=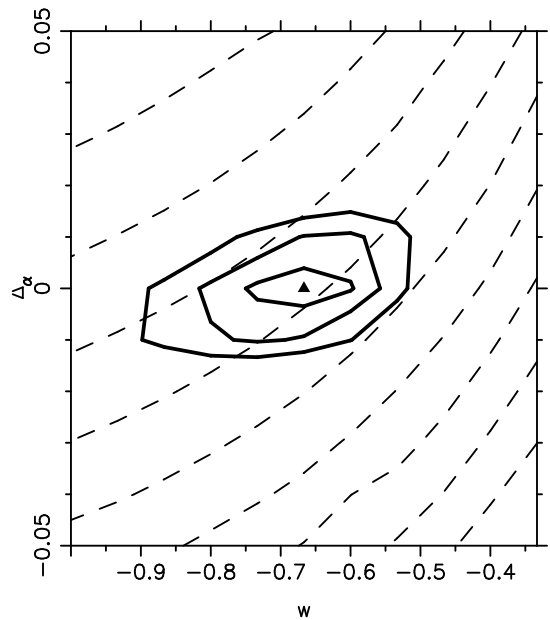,height=8.5cm}
\caption{ Likelihood contours in the $\left( w_{Q},
\Delta _{\alpha }\right)$
plane around a fiducial model (marked with a solid triangle)
with $w_{Q}=-2/3$, $\Delta _{\alpha }=0$. The models are compared
to the fiducial model at each $C_{\ell }$ out
to $\ell _{max}=1500$. The solid curves are
 $68.3\%$, $95.5\%$ and $99.7\%$. Likelikhood bounds,
and the dashed lines are contours of constant baryon density.
The fiducial model has $\Omega_B h^2 = 0.0200$
and $\Omega_B h^2$ changes by 0.0008
between the dashed contours.
 Note the great reduction in degeneracy
that occurs when one includes the damping scale into the comparison
of CMB anisotropy spectra.}
\label{fig_degen_qFM_1500}
\end{figure}

When searching for a degeneracy in the $(w_Q,\Delta_{\alpha})$ space, other
parameters were allowed to vary as well. Allowed ranges for $h$, $\Omega_B h^2$
and $n_s$ were $[0.5,0.9]$, $[0.015,0.025]$ and $[0.85,1.15]$ respectively.
At all grid points in the $(w_{Q},\Delta_{\alpha })$ plane the
best-fit values for $h$ and $n_{s}$ never varied by more than
$1\%$ from their fiducial values. Baryon density, by contrast,
changed by about $4\%$ in Fig.~\ref{fig_degen_qFM_900} for every $1\%$
change in the fine structure constant. The thin dashed lines in
Figs.~\ref{fig_degen_qFM_900} and~\ref{fig_degen_qFM_1500}
are contours of constant $\Omega _{B}h^{2}$, showing that
some variation of the baryon density is required, as well as $w_Q$,
to compensate for the effects of a change in the fine structure
constant. However, merely varying $\Omega_{B}h^{2}$, while holding
$w_Q$ fixed, is not sufficient to compensate for the change in $\Delta_{\alpha }$.
Figure~\ref{fig_baryoncontrol_qFM_900} shows the likelihood contours in the
$(\Omega_{B}h^{2},\Delta_{\alpha})$ space with the value of $w_{Q}$ fixed at $-2/3$.
These likelihood contours enclose a compact region, rather than an
elongated, curved region as in Fig.~\ref{fig_degen_qFM_900}.
That is, if $w_Q$ is not allowed to vary, the degeneracy is not present. Though
a change in the baryon density is necessary to compensate for a change
in $\Delta _{\alpha }$, because the change in redshift of the last-scattering
surface means the baryon density was different at the last-scattering,
a change in $w_{Q}$ is necessary to compensate for the change in the
angular scale of the sound horizon on the last-scattering surface.

The degeneracy does not extend down to the damping scale.
As shown in Fig.~\ref{fig_degen_qFM_1500}, the
degeneracy is broken if the measurement of the $C_{\ell}$'s
is extended to $\ell_{max}\sim 1500$ while still assuming a precision
of $5\%$ + cosmic variance. It is worth noting that when $\ell_{max}\sim 1500$
the spectra are very nearly identical from the 2nd acoustic peak up to and including the
damping scale. The degeneracy can not match the entire spectrum and it is the
smaller scales that carry the greater statistical weight. The degeneracy
is broken due to differences that arise at large scales (between $\ell=2$
and the first Doppler peak). It is worth noting, however, that we have only considered
the scalar contribution to the CMB anisotropy spectrum.
The addition of the tensor contribution would affect the spectrum predominantly
at large scales before the first Doppler peak, which is where the
spectra discussed above differ most. Thus, it is very
likely that with the addition of a tensor contribution, with a
spectral index $n_{t}$ and a relative normalization $r$ taken
as free parameters, the degeneracy would no longer be broken by the
damping scale. This is a subject of ongoing work.

Other degeneracies in CMB parameters, also involving quintessence,
were studied in Ref. \cite{Huey99}. Ways of breaking these
degeneracies by other cosmological observations were
discussed. In light of our current result, the extent to which, for example,
$w_Q$ can be resolved by combining multiple types of observations,
as discussed in~\cite{Huey99}, must be revisited.

\section{Outlook}
\label{outlook}

We have shown that the effect of a varying $\alpha$ on the CMBR
can be partially compensated by adjusting the value of the equation
of state of the dark energy. Namely, the value of $\alpha$ at recombination decreases with
the decrease in $w_Q$ along the degeneracy line. This limits the accuracy with which one
could determine \( \Delta_\alpha \) or the quintessence equation of state \( w_Q \)
from CMB observations alone.

There is a number of additional effects that could have been taken into account.
Some of them, such as the effect of a varying $\alpha$ on the
helium abundance, are relatively small. However, there can be
additional non-trivial effects if the dynamics and fluctuations in the field
which drives the change in $\alpha$ are also considered to a full extent.
A time-dependent $w_Q$ is yet another possibility that was not considered in this paper.

Both phenomena, varying $\alpha$ and quintessence, can be
modeled within Einstein's theory as a light scalar
field either minimally or non-minimally coupled to gravity \cite{Moffat,Bean00}.
These theories closely resemble dilaton and Brans-Dicke gravities.
It has also been argued that the same light scalar field which is responsible for
varying $\alpha$ could give rise to quintessence \cite{Dvali01}.
Other investigations have also pointed to a possible connection
between varying $\alpha$ and dark energy. Barrow, Sandvik and Magueijo have analyzed
the behavior of a varying $\alpha$ cosmology during the radiation, dust, curvature and
cosmological constant domination epochs \cite{Haavard}.
In their model the value of $\alpha$ increases during the matter domination
but rapidly approaches a constant when negative curvature or $\Lambda$ start to dominate.
A similar effect could have been achieved if a quintessence field was used in place
of a cosmological constant.

The degeneracy described in this paper suggests a need for a
firmer theoretical understanding of how varying $\alpha$ and quintessence may
be related to each other as already suggested by \cite{Dvali01}.
It could be that the standard computer codes, such as CMBFAST \cite{cmbfast},
for calculating the CMBR spectra will have to be modified to fully
include both effects assuming that they are rooted in same underlying microphysical
phenomenon. We leave this for future investigation.

\acknowledgements
We are grateful to Robert Caldwell for making his quintessence
code available to us. We thank Ian Del Antonio, Rachel Bean, Joao Magueijo and
Haavard Sandvik for helpful comments. We acknowledge the use of CMBFAST \cite{cmbfast}.
LP is supported by PPARC.

\end{document}